\documentclass[twocolumn]{article}
\usepackage{amsmath}
\usepackage{amssymb}
\begin{document}
\title{Pendulum limit, chaos and phase-locking in the dynamics of ac-driven 
 semiconductor superlattices}
\author{
Kirill N. Alekseev$^{1,2}$\thanks{E-mail: Kirill.Alekseev@oulu.fi},
Feodor V. Kusmartsev$^{3}$\thanks{E-mail: F.Kusmartsev@lboro.ac.uk}\\ 
$^{1}$Department of Physical Sciences, P.O. Box 3000, University of Oulu 
FIN-90014, Finland\\
$^{2}$Theory of Nonlinear Processes Laboratory, Kirensky Institute of
Physics, Krasnoyarsk 660036, Russia\\
$^{3}$Department of Physics, Loughborough University, Loughborough, LE11 3TU, UK
}
\date{}
\maketitle
\begin{abstract}
We describe a limiting case when nonlinear dynamics of an ac-driven 
semiconductor superlattice in the miniband transport regime is governed by a 
periodically forced and damped pendulum. We find analytically the conditions 
for a transition to chaos and consider an influence of temperature on the 
effect. We also discuss fractional dc voltage states in a superlattice 
originating from phase-locked states of the pendulum.
\end{abstract}

\medskip
\hrule
\bigskip

\section{Introduction}

Strongly nonlinear dynamical effects in miniband semiconductor
superlattices (SSLs) driven by an intense high-frequency electric field
attract growing interest in the last few years
\cite{ignatov95,alekseev96,alekseev98a,alekseev98b,%
alekseev99,cao00,romanov00,romanov01,alekseev01}. This theoretical 
activity, which includes investigation of dissipative chaos 
\cite{alekseev96,alekseev98b,cao00,romanov00}
and a spontaneous generation of dc voltage in a purely ac-driven SSL
\cite{ignatov95,alekseev98a,alekseev98b,alekseev99,romanov00,romanov01,%
alekseev01,alekseev02},
has been stimulated by a recent progress in the experimental studies of
nonlinear electron transport in the SSLs driven by THz field \cite{exp}.
On other hand, recent experiments \cite{exp1} reveal an important 
role of coherent plasma oscillations in nonlinear dynamics of carriers 
in SSLs.
\par
The counterparts of the effects of chaos and spontaneously dc voltage 
generation in the THz-driven SSL are
known in the rf-driven Josephson junctions for a rather long time
\cite{chaos-0,d'humieres,gubankov,langenberg,levinsen} (for a review, see 
\cite{kautz}).
So-called an inverse ac Josephson effect (a quantized dc voltage generation) 
\cite{langenberg,levinsen} has 
already found applications in the design of a zero-biased Josephson voltage 
standard \cite{kautz}, which offers several advantages in comparison with 
the conventional voltage standards based on an ac Josephson effect 
\cite{hamilton}. A transition to dissipative chaos often limits the stable 
operation of such kind of voltage standard \cite{kautz}. Nonlinear dynamics 
of driven plasma oscillations in a junction is well-described by
the fairly popular Resistively Shunted Junction (RSJ) model
\cite{mccumber-stewrt,likharev}, which is equivalent to a periodically forced 
and damped pendulum -- one of the most important paradigms in the modern 
nonlinear science \cite{sagdeev}.
\par
Within the semiclassical approach \cite{bass}, ac-driven plasma oscillations in
a single miniband SSL are described by the set of balance
equations for the average electron velocity (current), the average electron 
energy and the electric field (voltage) generated by an electron motion in SSL
\cite{ignatov95,alekseev96,alekseev98a,romanov00,romanov01,alekseev01}.
It has been shown in ref.~\cite{alekseev96} that this set of 
three nonautonomous nonlinear first-order
differential equations is isomorphic to the periodically forced Lorenz
model \cite{sparrow}. In comparison with the RSJ model, this system is more 
complex from a viewpoint of nonlinear dynamics and until recently it has been 
analysed mainly numerically 
\cite{ignatov95,alekseev96,alekseev98a,alekseev98b,romanov00,alekseev01}. 
\par
However, it should be noticed that a pendulum-like behavior of miniband
electrons in SSL has been established earlier for some limiting cases.
First, plasma oscillations in SSL have been described within a collisionless 
approximation by a pendulum without damping 
in \cite{epshtein,alekseev94,alekseev96}.
Second, describing the frequency-to-voltage conversion in ac-driven SSL,
Dunlap {\it et al.} \cite{dunlap} have found that their model of SSL 
demonstrates a behavior similar to the phase-locked solutions of damped 
pendulum \cite{langenberg} for $\omega\gtrsim\omega_{\rm pl}$ 
($\omega$ and $\omega_{\rm pl}$ are the ac electric field frequency and 
the miniband plasma frequency, correspondently). 
These earlier findings indicate that dynamics governed by a pendulum can be 
underlying for nonlinear dynamics of ac-driven SSL.
\par
In this work, we show that the driven damped pendulum could describe {\it
a stationary nonlinear dynamics} of an ac-driven SSL in the limit when a
characteristic scattering constant of electrons with impurities and phonons,
$\gamma$, is less than an an ac frequency $\omega$. 
In contrast to the RSJ model, where the voltage across junction is 
proportional to the velocity of pendulum, the voltage across SSL depends
on both the coordinate and the velocity of pendulum. These
findings possess us to apply several exact and approximate results obtained
earlier for a pendulum to {\it an analytic investigation} of regular 
and chaotic dynamics in SSL. Namely, we will show the following:\\
(1) Chaotic dynamics is natural for a relatively
low damping, i.e. for $\gamma/\omega_{\rm pl}\ll 1$.\\ 
(2) The SSLs with wide minibands are more stable against perturbations
leading to a transition to chaos in comparison to the case of narrow miniband 
SSLs.\\
(3) The Melnikov criterion of transition to chaos in the pendulum qualitatively
explains the position of low frequency boundary of transition to chaos in SSL.\\
(4) For a relatively low damping, $\gamma/\omega_{\rm pl}\ll 1$, 
phase-locking in driven pendulum results in a generation of dc bias across 
of SSL; the dc voltage per period of SSL is approximately proportional to a 
half-integer multiplier of $\hbar\omega$.\\ 
(5) The difference between generated dc voltage per superlattice's period and
$\hbar\omega$ is proportional to $(\gamma/\omega_{\rm pl})^2$ at 
$\gamma/\omega_{\rm pl}\ll 1$.\\
(6) For a fairly strong scattering, phase-locking in driven pendulum may 
not result in a generation of quantized (or an approximately quantized) dc bias 
across SSL.\\
Our present analytic research explains and further develops the results 
obtained numerically in the previous publications 
\cite{alekseev96,alekseev98a,alekseev98b,alekseev01}.

\section{Pendulum limit in superlattice balance equations}

We consider electron transport through a single miniband
of a spatially homogeneous SSL under the influence of ac current.
We assume the tight-binding dependence
$\varepsilon(k)=\Delta/2[1-\cos(k a)]$ of the electron energy
$\varepsilon$ on its quasimomentum $k$ along SSL's axis, where
$\Delta$ is the miniband width, and $a$ is the period of SSL.
The electron dynamics could be described by the following balance equations
\cite{ignatov95,alekseev96,alekseev98a,romanov00,romanov01,alekseev01}
(we use notations of ref. \cite{alekseev01})
\begin{eqnarray}
\dot{v}&=&u w -\gamma v,\nonumber\\
\dot{w}&=&-u v -\gamma (w-w_{\rm eq}),\label{balance}\\
\dot{u}&=&\omega_0^2 v+I_0\sin\omega t,\nonumber
\end{eqnarray}
Here the scaled variable $u=e a E(t)/\hbar$ describes the electric field $E$ 
(or the voltage per SSL's period), generated by an electron motion along the 
SSL axis;
$v=m_0 \overline{V} a/\hbar$ is the scaled electron velocity ($\overline{V}$ is
the electron velocity averaged over the time-dependent distribution function 
satisfying the Boltzmann equation,
$m_0=\frac{2\hbar^2}{\Delta a^2}$ is the effective electron mass
at the bottom of miniband) and
$w=(\overline{\varepsilon}-\Delta/2)(\Delta/2)^{-1}$, where
$\overline{\varepsilon}$ is the average electron energy within the miniband.
Following this scaling for the energy, the lower (upper) edge of the miniband
corresponds to $w=-1$ ($w=+1$). Next, $\omega$ is the frequency and $I_0$ is 
the scaled amplitude of driving ac current; $\gamma$ is the phenomenological 
relaxation constant describing different channels of electron scattering in SSL;
$\omega_0=\omega_{\rm pl}\left( C_{\rm sl}/C\right)^{1/2}$ is the generalized
plasma frequency, where $\omega_{\rm pl}=\left( 4\pi e^2 N/m_0\epsilon_0
\right)^{1/2}$ is the miniband plasma frequency, $N$ is the number of electrons
per unit volume, $\epsilon_0$ is an average dielectric constant for the
SSL material, $C_{\rm sl}=\frac{\epsilon_0 S}{4\pi l}$ is the self-capacitance 
of SSL with the length $l$ and the cross-section area $S$, and $C$ is 
a parasitic capacitance, which is in parallel to the SSL's capacitance.
The electron energy within miniband relaxes to
the thermal equilibrium energy value, $w_{\rm eq}$ ($w_{\rm eq}\leq 0$); 
in the case of a nondegenerate electron gas, $w_{\rm eq}$ has the following 
temperature dependence \cite{ignatov76,balance-ref}
\begin{equation}
\label{w0}
w_{\rm eq}=-\frac{I_1(\Delta/2k_BT)}{I_0(\Delta/2k_BT)},
\end{equation}
where $I_{0,1}$ are the modified Bessel functions, $T$ is the lattice 
temperature, and $k_B$ is the Boltzmann's constant. Note that the dependence 
(\ref{w0}) well describes temperature-induced modifications of the Drude 
conductivity in the narrow miniband SSLs as have been observed in the 
experiments \cite{brozak,sibille}.
\par
First two equations of the set (\ref{balance}) are the well-known balance 
equations of Ignatov and Romanov \cite{ignatov76,balance-ref},
which could be derived from the Boltzmann transport equation with a constant
relaxation time $\gamma$.
Thus, instead of finding the time-dependence of distribution function $f(p,t)$
$(p=\hbar k)$ from the Boltzmann equation and further calculation of
the average electron velocity $\overline{V}=\int dp f(p,t)\partial
\varepsilon(p)/\partial p$ and the 
average electron energy $\overline{\varepsilon}=\int dp f(p,t) \varepsilon(p)$,
it is possible just to solve the corresponding balance equations for the 
average variables. Both these approaches give the same result for the
tight-binding dispersion relation. 
\par
The third equation in (\ref{balance}) could be obtained from the Maxwell or 
Kirchoff equations \cite{ignatov95,alekseev96,alekseev98a,romanov00,romanov01}.
To the best our knowledge, three coupled balance equations 
in the form (\ref{balance}) have appeared first in the work of Tetervov
\cite{tetervov}, who used these equations for a description of decay of 
undriven ($I_0=0$) nonlinear plasma oscillations in SSL.
\par
Now we want to show how the pendulum limit arises in the superlattice 
balance equations (\ref{balance}).  We introduce new variables $A$ and $\theta$
as $w=-A\cos\theta$, $v=-A\sin\theta$. We can re-write Eqs. (\ref{balance})
in the form (see Appendix \ref{sec:app-A})
\begin{eqnarray}
\ddot{\theta}&+&A\omega_0^2\sin\theta+\gamma G(\theta,A)\dot{\theta}=
\nonumber \\
& & I_0\sin\omega t + \gamma^2w_{\rm eq} A^{-1} F(\theta,A), 
\label{complex_pend}\\
\dot{A}&=&-\gamma A-\gamma w_{\rm eq}\cos\theta,\label{complex_A}
\end{eqnarray}
where
\begin{eqnarray}
G(\theta,A)&=&-w_{\rm eq} A^{-1}\cos\theta,\label{complex_G}\\
F(\theta,A)&=&\sin\theta+w_{\rm eq} A^{-1} \sin\theta\cos\theta,
\label{complex_F}\\
\dot{\theta}(t)&=&u(t)+\gamma w_{\rm eq} A^{-1}(t)\sin\theta(t)
\label{complex_field}.
\end{eqnarray}
Formally, the set of equations (\ref{complex_pend})-(\ref{complex_field})
is not simpler than the original balance equations (\ref{balance}). However,
this form is more suitable for the consideration of pendulum limits. 
\par
We start our analysis with the case of collisionless SSL, $\gamma=0$. 
As follows from  Eq. (\ref{complex_A}), $A^2=v^2(t)+w^2(t)$ is the integral 
of motion in this case. For natural 
initial conditions $v(0)=0$, $w(0)=w_{\rm eq}$, we have $A=-w_{\rm eq}$.
The balance equations are reduced to single equation of 
undamped and periodically forced pendulum 
\begin{equation}
\label{pend_ham}
\ddot{\theta}+(-w_{\rm eq})\omega_0^2\sin\theta=I_0\sin\omega t,
\end{equation}
where $v(t)=w_{\rm eq}\sin\theta(t)$, $w(t)=w_{\rm eq}\cos\theta(t)$,
$u(t)=\dot{\theta}(t)$. Earlier the same equation has been obtained
within the semiclassical Hamiltonian approach by several authors
\cite{epshtein,alekseev94}. It was used in the investigations of nonlinear
plasma oscillations \cite{epshtein}, the instabilities and the
Hamiltonian chaos \cite{alekseev94} in ideal ac-driven SSLs.
\par
Eq.~(\ref{pend_ham}) is valid only for a short time $t\ll\gamma^{-1}$. Now 
we want to show that a pendulum equation can also describe
electron's dynamics in the ac-driven SSL at long time ($t\gg\gamma^{-1}$)
providing $\omega\gg\gamma$.
Eq.~(\ref{complex_pend}) represents a pendulum, parameters of which are 
varied self-consistently due to their dependences on $A$ and $\theta$. 
To begin with we limit our 
attention to the time interval $T=2\pi/\omega$. During this characteristic 
period of oscillations (or rotations), $T$, the variable $A$ only lightly 
decays for $\omega\gg\gamma$ (see Eq.~(\ref{complex_A})). Now we compare 
characteristic damping terms in Eqs.~(\ref{complex_pend}) and (\ref{complex_A}).
Following Eqs.~(\ref{complex_pend}) and (\ref{complex_G}), the averaged over 
period of ac field dissipative term is 
$$
\langle\gamma G(\theta,A)\rangle=-\gamma w_{\rm eq}\langle A^{-1}\cos\theta
\rangle
$$
Thus, the ratio of characteristic dissipative terms in 
Eqs.~(\ref{complex_pend}) and in Eq.~(\ref{complex_A}),
calculated during the time interval $T$, is following
$$
\frac{\langle\gamma G(\theta,A)\rangle}{\gamma}=-w_{\rm eq}\langle A^{-1}
\cos\theta\rangle\ll 1
$$
This ratio is small because the term $\cos\theta$ changes its signs 
(oscillates), while $A$ stays almost constant. Next,
we consider more long time interval of more number of periods $T$ and come to
a conclusion that a dissipation in Eq.~(\ref{complex_A}) is, in effect, much 
more strong than a dissipation in Eq.~(\ref{complex_pend});
the variable $A(t)$ approach its stationary value faster than the variable 
$\theta(t)$ approaches its attractor.
Therefore, if additionally the amplitude of driving current is not large,
$I_0/\omega^2_0<1$, the variable $A$ can be adiabatically eliminated 
\cite{haken}. That is, we can substitute the stationary value 
$A(t)\rightarrow A_{\rm st}(t)=-w_{\rm eq}\cos\theta$ in 
Eq.~(\ref{complex_pend}). Observing that $G(\theta,A_{\rm st})=1$ and
$F(\theta,A_{\rm st})=0$, we find that Eqs.~(\ref{complex_pend}), 
(\ref{complex_field}) take the form
\begin{equation}
\label{pend}
\ddot{\psi}+\gamma\dot{\psi}+(-w_{\rm eq})\omega_0^2\sin\psi=2 I_0\sin\omega t,
\end{equation}
\begin{equation}
\label{field}
u=\dot{\psi}/2+\gamma\tan(\psi/2),
\end{equation}
where we introduced $\psi(t)=2\theta(t)$. This pendulum representation
of superlattice balance equations constitutes the main result of this paper.
\par
It is useful to compare some well-known solution of 
Boltzmann equation for the tight-binding lattice and the corresponding 
solution of pendulum equation (\ref{pend}).
This test can be simplified employing an analogy with the RSJ model 
of Josephson junctions, which is also governed by a periodically driven 
and damped pendulum. Consider a SSL driven by a given ac electric field 
$E=E_0\cos\omega t$ with the frequency being much greater than the miniband 
plasma frequency, $\omega\gg\omega_0$. In this case we have 
$I_0=\omega_s\omega$ with $\omega_s=e E_0 a/\hbar$ 
\cite{alekseev96,romanov00,romanov01,alekseev01}.
For such $I_0$ and $\omega$, we find that the solution of pendulum equation 
for the variable $v(t)=\sin\theta\cos\theta$ is same as the well-known 
expression for the time-dependent electron's velocity $v(t)$ derived from
the exact solution of the Boltzmann equation \cite{ignatov76,tsu71}, if 
$\omega\gg\gamma$ and $I_0/\omega^2_0<1$ (see Appendix \ref{sec:app-R}). 
Therefore, we see that the pendulum representation, 
Eqs.~(\ref{pend})-(\ref{field}), gives same result
as the Boltzmann equation within the range of its validity.
Moreover, this range of parameters, $I_0/\omega^2_0<1$ and $\omega\gg\omega_0$, 
corresponds to $z\equiv\omega_s/\omega\ll 1$ for $\omega\gg\omega_0$,
what are quite typical values of the field strengths and the frequencies in the 
modern experiments on harmonics generation and detection of THz radiation in 
SSLs \cite{exp}.
Finally, it is worth to notice that from the viewpoint of analogy
with the Josephson junctions, a described generation of harmonics of an ac 
electric field in SSL
is completely equivalent to the standard ac Josephson effect 
\cite{kautz,likharev},
i.e. to a generation of current's harmonics under the action of a sinosuidal
voltage.
\par
In this paper we are mainly focused on nonlinear dynamical effects arising
within the pendulum limit.
With the reference to the pendulum representation of balance equations, we can 
make the important conclusion: {\em The thresholds of instabilities in 
an ac-driven semiconductor superlattices, including the transition to chaos, 
are same as in the RSJ model of an ac-driven Josephson junction providing} 
$\omega\gg\gamma$. 
\par
Before we proceed with a consideration of instabilities and chaos,
we should note that the SSL balance equations 
(\ref{complex_pend})-(\ref{complex_field})
can be also simplified in the opposite limiting case of frequent collisions, 
$\omega\ll\gamma$. Really, assuming $|\dot{\theta}|\simeq\omega\ll\gamma$ 
we see from Eq. (\ref{complex_A}) that $A(t)$ follows $\theta(t)$ 
adiabatically. Therefore Eqs. (\ref{complex_pend}), (\ref{complex_G}), 
(\ref{complex_F}), (\ref{complex_field}) take the form
\begin{equation}
\label{overdamped-pend}
\gamma\dot{\psi}+\omega_0^2\sin\psi=2 I_0\sin\omega t
\end{equation}
with $u=\gamma\tan(\psi/2)$. 
Such kind of the first order equation is well-known in the theory of Josephson
junctions \cite{aslamazov,likharev}. Importantly, it also arises in the 
description of an interaction of electromagnetic radiation with the lateral 
SSLs \cite{alekseev02,lssl}.
Chaos is impossible in the first order differential equation \cite{sagdeev}. 
However, the overdamped pendulum (\ref{overdamped-pend}) driven by 
a two-frequency quasiperiodic force still can demonstrate very complex dynamics 
known as a strange nonchaotic attractor \cite{sna}.
\par
Now we turn to the consideration of conditions for a transition to chaos and 
an influence of temperature effects
using the pendulum representation of balance equations (\ref{pend}),
(\ref{field}).

\section{Periodicity and chaos}

The theorem of Levi \cite{levi} states that the stationary dynamics of 
a periodically forced pendulum is never chaotic in the overdamped limit, which 
in our case reads
\begin{equation}
C\equiv\frac{\gamma}{|w_{\rm eq}|^{1/2}\omega_0}>2.
\label{overdamped}
\end{equation}
The numerical simulations of the driven damped pendulum performed for the wide 
ranges of $I_0$ and $\omega$ reveal the absence of chaos already for 
$C\approx 1$ \cite{taiwan}.
\par
For a narrow miniband at room temperature, the thermal equilibrium energy 
becomes close to the center of miniband, i.e. $|w_{\rm eq}|$ is small 
(see Eq. (\ref{w0})). Therefore, {\it the criterion for absence of chaos 
(\ref{overdamped}) can be easier satisfied in the SSLs with
narrow miniband in a comparison to the case of wide miniband SSLs}. 
In particular, for $k_B T\gg\Delta$ we have $w_{\rm eq}\approx -\Delta/2 k_B T$ 
and therefore $C\propto\gamma/\Delta$ (really, $\omega_0\propto m_0^{1/2}
\propto\Delta^{1/2}$ providing $|w_{\rm eq}|^{1/2}\omega_0\propto\Delta$).
Thus, with a decrease of miniband width $\Delta$, the value of $C\propto\gamma
/\Delta$ increases making the condition (\ref{overdamped}) to be more robust.
\par
As an example we consider the narrow miniband SSL with $\Delta=3$ meV, 
$a=22$ nm, $N=2\times 10^{15}$ cm$^{-3}$ \cite{brozak} and 
$\gamma^{-1}\approx 0.5$ ps \cite{sibille}, the miniband plasma
frequency is $\omega_{\rm pl}=2.1\times 10^{12}$ rad/sec and 
$\gamma/\omega_{\rm pl}\approx 1$. For the temperature $T=50$ K, $w_{\rm eq}
\approx -0.33$ and the factor $|w_{\rm eq}^{-1/2}|$ is 1.7. For the higher 
temperature, $T=200$ K, $|w_{\rm eq}^{-1/2}|\approx 3.5$ and chaos is always 
impossible. However, chaos is still possible for a wider miniband and a longer 
relaxation time (see corresponding estimates in 
Refs.~\cite{alekseev96,alekseev98b,romanov00,alekseev01}).
\par
The criterion of absence of strange attractor, Eq.~(\ref{overdamped}), has 
been obtained within the pendulum limit and thus is formally valid only for 
$\omega\gg\gamma$. 
However, summing up the results of numerical simulations of the superlattice 
balance equations, we speculate that strange attractor does not exist at
any frequency $\omega$, if the inequality (\ref{overdamped}) is satisfied.
\par
To find an analytic criterion of a transition to chaos in the pendulum limit,
we apply the method of Melnikov \cite{mel'nikov}.
This method has been used by several authors for a determination of the 
conditions of transition to chaos in the periodically driven pendulum 
(or for the RSJ model) \cite{gubankov,kautz}. In our case, the Melnikov's 
criterion for a transition to chaos is 
\begin{equation}
I_0>I_0^{\rm cr}=\frac{2\gamma\omega_0 |w_{\rm eq}|^{1/2}}{\pi}\cosh
\left(\frac{\pi}{2}\frac{\omega}{|w_{\rm eq}|^{1/2}\omega_0}\right)
\label{chaos-melnikov}
\end{equation}
(cf. Eq. (60) in Ref. \cite{kautz}).
This formula is derived with the assumption that conditions 
$$
\frac{\gamma}{|w_{\rm eq}|^{1/2}\omega_0}\ll 1, \quad 
\frac{2 I_0}{|w_{\rm eq}|\omega_0^2}\ll 1
$$
are satisfied. For $\omega<\omega_0$ Eq. (\ref{chaos-melnikov}) takes the form
\begin{equation}
I_0^{\rm cr}=\frac{2}{\pi}\gamma\omega_0 |w_{\rm eq}|^{1/2},
\label{chaos-melnikov1}
\end{equation}
In the case of SSL driven by the ac electric field $E=E_0\cos\omega t$, 
one can get $I_0=\omega_s\omega$ with $\omega_s=e E_0 a/\hbar$ 
\cite{alekseev96,romanov00,romanov01,alekseev01}.
Then, the formula (\ref{chaos-melnikov1}) gives for the boundary of chaotic
region in the $\omega_s$-$\omega$ plane in the form
\begin{equation}
\omega_s\omega=\frac{2}{\pi}\gamma\omega_0 |w_{\rm eq}|^{1/2}.
\label{chaos-melnikov2}
\end{equation}
Earlier, the low-frequency boundary of chaos in the simple form 
$\omega_s\omega=\mbox{const}$ has been found numerically in the works 
\cite{alekseev96,alekseev98b,dissert},
but has been unexplained until now.
Eq. (\ref{chaos-melnikov2}) can qualitatively explain this dependence.

\section{Phase-locking in pendulum and dc voltage generation in superlattice}
\label{sec-phase-lock}

Phase-locking is another nonlinear dynamic phenomenon which an ac-driven 
pendulum can demonstrate along with chaos \cite{sagdeev}. 
In the physics of Josephson junctions, phase-locking at zero bias is
known as the inverse ac Josephson effect 
\cite{langenberg,levinsen,kautz,hamilton}. Here we are 
discussing the influence of phase-locking in the pendulum on the physical 
properties of SSL. In the driven pendulum phase-locking means that
\begin{equation}
\langle\dot{\psi}\rangle=\frac{n}{l}\omega,
\label{phase-locking}
\end{equation}
where $n$ and $l$ are integer numbers
and $\langle\ldots\rangle$ stands for the time averaging over the period of ac 
drive $2\pi/\omega$ \cite{d'humieres,kautz}. Majority of stable 
phase-locked states are integer ($n\neq 0$, $l=1$); however, a fractional 
phase-locking ($l>1$) can also exist. In the case of Josephson junction, the
voltage $U$ across the junction is proportional to $\dot{\psi}$; therefore 
phase-locking results in a generation of quantized dc voltage 
$\langle U\rangle\propto (n/l)\omega$ \cite{levinsen,kautz,likharev}. 
\par
For the superlattice problem, the voltage is a function
of both the velocity and the phase of the pendulum (see Eq. (\ref{field})). 
Therefore, phase-locking in the pendulum (\ref{phase-locking}) 
determines a generation of dc voltage in the SSL as
\begin{eqnarray}
\langle u\rangle &=& \frac{n}{l}\frac{\omega}{2}\left( 1+
\frac{\gamma}{\omega_0} R_{nl}\right),\label{dc-field} \\
R_{nl}&=&\frac{2\omega_0}{\omega}\frac{l}{n}\langle\tan\left(
\frac{\psi}{2}\right)\rangle.\label{R-def}
\end{eqnarray}
For a low damping $\gamma/\omega_{\rm pl}\ll 1$, we immedeatly get from 
(\ref{dc-field}) the dc voltage $\langle u\rangle\approx (n/2l)\omega$. 
Such kind of almost {\it half-integer} dc voltage states in a pure ac-driven 
SSL have been recently observed in the numerical simulations of balance 
equations \cite{romanov00,alekseev01}. 
\par
The formula (\ref{dc-field}) tells us nothing about a stability of dc 
voltage states in SSL that correspond to the phase-locked states of the 
pendulum with different $n$ and $l$. However, numerical simulations demonstrate 
that the dc voltage states, which are close to the integer states (i.e. to 
the states with $l=1$ and $n$ being {\it even integer}), are more typical then 
fractional states \cite{alekseev98a,alekseev01}.
\par
What is important, the dc voltage per superlattice period is not exactly
$(n/2l)\omega$ even for a weak damping. Moreover, a weak dependences of 
$\langle u\rangle$ on the ac amplitude \cite{alekseev98b,dissert} and the
frequency \cite{ignatov95,dissert} have been found numerically at
$\gamma/\omega_{\rm pl}\ll 1$. 
The appearance of such kind of dependences within the pendulum limit can be 
understood, if one takes into an account
that $\langle u\rangle$ is a function of both the time-average velocity and 
of the time-average coordinate of pendulum (see Eq. (\ref{dc-field})). 
Phase-locked rotational state of the pendulum can be represented in the form 
\cite{pedersen,d'humieres}
\begin{equation}
\psi(t)=\psi_0+n\omega+\sum_{p=1}^{\infty}\alpha_p \sin(p\omega t+\mu_p),
\label{phl-state-general}
\end{equation}
where $\psi_0$, $\alpha_p$ and $\mu_p$ are constants and we restrict our
consideration to the leading phase-locked states, $\langle\dot{\psi}\rangle=
n\omega$. The amplitudes $\alpha_p$ and the phases $\mu_p$ in 
Eq.~(\ref{phl-state-general}) are functions of the main physical parameters 
of our problem, i.e. $I_0$, $\omega$ 
and $\gamma$. Substituting the expression (\ref{phl-state-general}) in 
Eq. (\ref{R-def}) one can see that the dependence of dc voltage 
$\langle u\rangle$ on the ac amplitude and frequency arise in 
Eq. \ref{dc-field} via the term $R_{nl}$.
\par
It is important to know how this ``dissipative correction'' to the quantized dc
voltage, $\langle u\rangle-n\omega/2=(\gamma/\omega_0) R_n$, scales with an
increase of dissipation. To estimate the dependence of $R_n$ on $\gamma/
\omega_0$ for several lowest phase-locked states of the pendulum,  we 
consider the limit of fast rotations $\alpha\ll 1$. In this case formula
(\ref{phl-state-general}) takes the form  
\begin{equation}
\psi(t)=\psi_0+n\omega+\alpha\sin(\omega t+\mu).
\label{phl-state}
\end{equation}
Substituting Eq. (\ref{phl-state}) in Eq. (\ref{pend}) and equating zero 
harmonics, we have the formula (cf. \cite{d'humieres,pedersen}) 
\begin{equation}
\omega_0^2 J_n(\alpha)\sin(n\mu)=-n\gamma\omega,
\label{0-harmonic}
\end{equation}
which gives us an opportunity to understand the characteristic dependences of 
$\alpha$ and $\psi$ on $\gamma$ and $\omega$. Next, substituting 
Eq. (\ref{phl-state}) in Eq. (\ref{R-def}) and calculating integrals
in the limit $\alpha\ll 1$, we obtain (see Appendix \ref{sec:app-B})
\begin{eqnarray}
\langle\tan\left(\psi/2\right)\rangle_{n=1}&=&
\alpha\sin(\mu)+{\cal O}(\alpha^2),\label{R-finding}\\
\langle\tan\left(\psi/2\right)\rangle_{n=2}&=&
-0.25\alpha^2\sin(2\mu)+{\cal O}(\alpha^4). \nonumber
\end{eqnarray}
Now combining Eqs.~(\ref{R-def}), (\ref{0-harmonic}), 
(\ref{R-finding}) and using the asymptotics of the Bessel
functions $J_n(z)\approx z^n/(2^n n!)$ ($z\ll 1$), we find
\begin{equation}
R_n\simeq\left( -1 \right)^n \frac{4\gamma}{\omega_0},\quad\mbox{(n=1,2)}.
\label{R-result}
\end{equation}
Thus, for a weak dissipation ($\gamma/\omega_{\rm pl}\ll 1$), the dc voltage 
spontaneously generated in a pure ac-driven SSL is
\begin{equation}
\langle u\rangle=n\omega/2+{\cal O}(\gamma^2/\omega^2_{\rm pl})
\label{voltage-final}
\end{equation}
and it has a quadratic dependence on $\gamma$.
\par
With a further increase of $\gamma$, the dissipative term in 
Eq. (\ref{dc-field}) becomes more important. Moreover, for a large 
dissipation the phase-locked rotational states (\ref{phl-state-general}), 
(\ref{phl-state}) do not exist in the pendulum anymore \cite{d'humieres}.
However, the unquantized dc voltage still can be generated in SSL due to 
different mechanisms, as has been demonstrated numerically in \cite{alekseev01}.
The discussion of relationship between a pendulum dynamics and a dc voltage 
generation in a superlattice for an arbitrary damping will be presented 
elsewhere.

\section{Discussion and conclusion}

It is instructive to compare the pendulum equation (\ref{pend}), derived from
the superlattice balance equations in the limit of rare collisions ($\gamma\ll
\omega$) and corresponding motion equation that can be obtained 
employing the Newton law. We introduce  simplest form of phenomenological 
dissipative term to the motion equation for electron's momentum as
$$
\dot{p}+\gamma p =-e E(t).
$$
Further, combining this equation and the standard definition of electron's 
velocity $V=\varepsilon(p)/\partial p$, the tight-binding dispersion relation,
and the Maxwell equation (third equation in the set (\ref{balance})), 
we get the pendulum equation in the form
$$
\ddot{\phi}+\gamma\dot{\phi}+\omega_0^2\sin\phi=I_0\sin\omega t,
$$
where $\phi=p a/\hbar$, $|w_{\rm eq}|\equiv 1$ for simplicity. In this 
approach the scaled voltage is $u=\dot{\phi}$.
Certainly, this equation is similar to Eq. (\ref{pend}). The difference arises
mainly in the dependence of voltage on coordinate and momentum of the pendulum
(cf. Eq. (\ref{field})). In particular, the voltage is dependent not only on
the momentum of pendulum but also on its coordinate in the Boltzmann equation
approach.
\par
The crude approach based on the Newton law uses a trajectory of an
individual electron. Therefore, this approach can not take into an account the
changes in a distribution function of electrons induced by an electric field.
In contrast, the approach based on the Boltzmann equation does include such 
kind of corrections in the term proportional to $\gamma$. In the limit of
infrequent collisions, $\gamma\ll\omega$, this term can be often ignored 
because $\dot{\theta}\simeq\omega\gg\gamma$. However, the dissipative 
contribution to voltage can become important if one considers a dc voltage
generation, $\langle u\rangle\neq 0$. As we have demonstrated in 
section~\ref{sec-phase-lock}, this dissipative correction can be important
in the consideration of phase locking, $\langle\dot{\theta}\rangle=\omega$.
The contribution of dissipative term to the dc voltage, $\langle u\rangle$, 
can be even more important in the case of lateral superlattices, for which
$\langle\dot{\theta}\rangle=0$ \cite{alekseev02}.
\par
In summary, we show that pendulum equation can well describe
the dynamics of electrons in a miniband of ac-driven semiconductor superlattice
in the limit of weak scattering. We demonstrate that several earlier numerical 
findings concerning chaos and spontaneous dc voltage generation in the 
superlattices can be explained within such approach.
We establish a link between the theory of diffusive transport in 
ac-driven superlattices and the theory rf-driven Josephson junctions. 
We point out that dissipative terms arising in the dependence of voltage 
across superlattice on the momentum and the coordinate of the pendulum should 
play an important role in many situations.
\par
We believe that described in this work simple and instructive theory can
be useful in an explanation of the results of experiments \cite{exp,exp1}
as well as that it can stimulate other experiments on the THz-field induced 
nonlinear phenomena in superlattices.

\section*{Acknowledgements}
We are grateful to David Campbell, Ethan Cannon, Alexander Zharov, Pekka 
Pietil\"{a}inen for collaboration. We thank Laurence Eaves, Martin Holthaus, 
Anatoly Ignatov, Karl Renk, Mikko Saarela, Erkki Thuneberg for discussions.
This research was partially supported by the Academy of Finland and the
Royal Society.

\appendix
\section{\label{sec:app-A}Appendix.}

In this appendix we will derive the equations
(\ref{complex_pend}), (\ref{complex_A}), (\ref{complex_G}),
(\ref{complex_F}), (\ref{complex_field}). To begin with we introduce a
complex variable $Z=w+i v$ and represent the superlattice balance equations
(\ref{balance}) in the form
\begin{equation}
\dot{Z}=-i u Z-\gamma Z +\gamma  w_{\rm eq}. 
\label{A1}
\end{equation}
Substituting the parametrization $Z=-A\exp(i\theta)$ 
(i.e. $w=-A\cos\theta$ and $v=-A\sin\theta$) in (\ref{A1}) and equating 
real and imaginary parts, we get
\begin{eqnarray}
\dot{A}&=&-\gamma A-\gamma w_{\rm eq}\cos\theta,\label{A2}\\
\dot{\theta}&=&u+\gamma w_{\rm eq} A^{-1}\sin\theta.\label{A3}
\end{eqnarray}
From the last equation we have
\begin{equation}
\label{A4}
\ddot{\theta}=\dot{u}+\gamma w_{\rm eq}
\frac{d}{dt}\left[ A^{-1}\sin\theta\right].
\end{equation}
Taking into account that
\begin{eqnarray}
& &\frac{d}{dt}\langle[ A^{-1}\sin\theta\rangle] =
-A^{-2}\dot{A}\sin\theta+A^{-1}\dot{\theta}\cos\theta \label{A5}\\
& &=\gamma A^{-1}\sin\theta+\gamma  w_{\rm eq} A^{-2}\sin\theta\cos\theta
+\dot{\theta} A^{-1}\cos\theta\nonumber
\end{eqnarray}
and using the third equation of set (\ref{balance}), we immediately get from 
Eq. (\ref{A4}) the Eqs. (\ref{complex_pend}), (\ref{complex_G}), 
(\ref{complex_F}). Finally, it is easy to see that Eqs. (\ref{A2}) and 
(\ref{A3}) are same as Eqs. (\ref{complex_A}) and (\ref{complex_field}), 
correspondently.

\section{\label{sec:app-R}Appendix.}

In this appendix we will show that well-known 
result of Tsu-Esaki-Ignatov-Romanov from the theory of ac-driven SSLs
\cite{ignatov76,tsu71} can be reproduced within the pendulum representation 
of the SSL balance equations (\ref{pend}), (\ref{field}).
\par
Solving the Boltzmann equation for the tight-binding lattice driven by 
the electric field $E=E_0\cos\omega t$ with $\omega\gg\omega_0$, 
Ignatov and Romanov \cite{ignatov76} have found for $\gamma\ll\omega$ 
the following time-dependence of the averaged electron's velocity 
\begin{equation}
\label{R1}
v(t)=\sum_{k=1}^{\infty} a_{2k-1}\sin(2k-1)\omega t, \quad 
a_{2k-1}=2 J_0(z) J_{2k-1}(z),
\end{equation}
where $z=\omega_s/\omega=(e E_0 a)/\hbar\omega$. Same result 
has been obtained earlier by Tsu and Esaki for the limiting case $z\ll 1$ 
\cite{tsu71}.
\par
Now turn to the pendulum equation. Hence we have the ac field of a very high 
frequency, we can neglect the last term ($\propto\gamma$) in 
Eq.~(\ref{field}) in comparison with the first term 
($\propto\dot{\theta}\simeq\omega$). Further, we can employ an analogy with the 
RSJ model of Josephson junctions. It is known for the RSJ model that the 
sinusoidal voltage is the solution of pendulum equation if 
$\omega\gg\omega_0$ and $\gamma\ll\omega$ \cite{kautz}. 
In our case it means that $\theta(t)=z\sin\omega t$ is the solution of 
Eq.~(\ref{pend}). Further, substituting this time dependence of $\theta(t)$
in the expression for averaged electron's velocity $v=\sin\theta\cos\theta$,
we can determine the current generated in the SSL under the action of the ac 
field.  
(Of course, it is the analog of the ac Josephson current \cite{kautz}).
Using the Bessel expansions
$$
\cos(z\sin t)=J_0(z) +2\sum_{k=1}^{\infty}J_{2k}(z)\cos 2kt,
$$
$$
\sin(z\sin t=2\sum_{k=1}^{\infty}J_{2k-1}(z)\cos(2k-1)t,
$$
we find the averaged electron's velocity in the form
\begin{eqnarray}
& & v(t)=2 J_0(z) \sum_{k=1}^{\infty}J_{2k-1}(z)\sin(2k-1)\omega t\label{R2}\\
& &+2\sum_{k_1,k_2\geq 1} J_{2k_1}(z) J_{2k_2-1}(z)\times\nonumber\\
& & \left[ \sin\left( 2(k_2-k_1)-1\right)\omega t +
\sin\left( 2(k_2+k_1)-1\right) \omega t \right]. \nonumber
\end{eqnarray}
Supposing that
\begin{equation}
\label{R3}
\omega_s<\frac{\omega^2_0}{\omega},
\end{equation}
we have
$$
z\equiv\frac{\omega_s}{\omega}<\frac{\omega^2_0}{\omega^2}\ll 1
$$
because $\omega/\omega_0\gg 1$. Now it is easy to see that for $z\ll 1$
Eq.~(\ref{R2}) is same as Eq.~(\ref{R1}).
\par
Thus, we have showed that for $\gamma\ll\omega$ and, additionally, if
the condition (\ref{R3}) is satisfied, the pendulum equation can reproduce
the result of Tsu-Esaki-Ignatov-Romanov. Finally, combining the definition
$I_0=\omega_s\omega$ and  Eq.~(\ref{R3}), we see that 
the inequality (\ref{R3}) just gives another condition of the
validity of the pendulum approach: $I_0/\omega^2_0<1$.

\section{\label{sec:app-B}Appendix.}

In this appendix we will derive the equations (\ref{R-finding}). We need to 
calculate the integrals
\begin{equation}
\label{B1}
\langle\tan (\psi/2)\rangle\equiv\frac{1}{2\pi}\int_0^{2\pi} dx 
\tan\left( \frac{n x}{2}+\frac{\alpha}{2}\sin(x+\mu)\right),
\end{equation}
where $n=1,2$. In the limit $\alpha\ll 1$ we can use the expansions
\begin{eqnarray}
\label{B2}
& &\tan\left( \frac{x}{2}+\frac{\alpha}{2}\sin(x+\mu)\right)=
\tan\left(\frac{x}{2}\right)+\nonumber \\
& &{\frac{\sin(x+\mu)}{\cos(x)+1}}\alpha-
{\cal O}\left({\alpha}^{2}\right), 
\end{eqnarray}
\begin{eqnarray}
& &\tan\left( x+\frac{\alpha}{2}\sin(x+\mu)\right)=
\tan(x)+\frac{\alpha}{2}\frac{\sin(x+\mu)}{\cos^2(x)}
\nonumber \\
& &-\frac{\alpha^2}{4} \frac{\left(-1+\cos^{2}(x+\mu)\right )
\sin(x)}{\cos^{3}x}+\nonumber \\
& & \frac{\alpha^3}{24}\frac{\left (-1+
\cos^2(x+\mu)\right)\sin(x+\mu)
\left (2\,\cos^{2}x-3\right)}{\cos^4 x}\nonumber \\
& & +{\cal O}\left({\alpha}^{4}\right).\label{B3}
\end{eqnarray}
After integration of the expression (\ref{B2}) and (\ref{B3}) we obtain Eqs.
(\ref{R-finding}).

\end{document}